\documentclass[letterpaper, 10 pt, conference]{ieeeconf}  

\IEEEoverridecommandlockouts                              

\usepackage{graphicx} 
\usepackage{epsfig} 
\usepackage{amsmath} 

\title{\LARGE \bf
MEMD-HHT based Emotion Detection from EEG using 3D CNN
}

\author{Monira Islam, Tan Lee
\thanks{Department of Electronic Engineering, The Chinese University of Hong Kong, Sha Tin, New territory, Hong Kong
        {\tt\small (email:1155131474@link.cuhk.edu.hk, tanlee@ee.cuhk.edu.hk)}}%
}

\begin{document}

\maketitle
\thispagestyle{empty}
\pagestyle{empty}

\begin{abstract}

In this study, the Multivariate Empirical Mode Decomposition (MEMD) is applied to multichannel EEG to obtain scale-aligned intrinsic mode functions (IMFs) as input features for emotion detection. The IMFs capture local signal variation related to emotion changes. Among the extracted IMFs, the high oscillatory ones are found to be significant for the intended task. The Marginal Hilbert spectrum (MHS) is computed from the selected IMFs. A 3D convolutional neural network (CNN) is adopted to perform emotion detection with spatial-temporal-spectral feature representations that are constructed by stacking the multi-channel MHS over consecutive signal segments. The proposed approach is evaluated on the publicly available DEAP database. On binary classification of valence and arousal level (high versus low), the attained accuracies are 89.25\% and 86.23\% respectively, which significantly outperform previously reported systems with 2D CNN and/or conventional temporal and spectral features.
\newline

\end{abstract}

\section{INTRODUCTION}

Emotion is an integral part of human cognitive processes and social behaviours. It is related with the mood, behaviour, thinking and preference of an individual. Emotional state is commonly described by a two-dimensional model, which comprises the valence and arousal levels \cite{c1}.
It has been shown that Electroencephalogram (EEG) based functional connectivity pattern analysis estimates the functional brain activity to detect the changes in emotional states \cite{c2}. The present study tackles the problem of binary classification of valence and arousal state, namely low valence vs. high valence, and low arousal vs. high arousal, using multi-channel EEG. The valence level is related to the positive or negative state of emotion, and the arousal level indicates intensity of the associated emotion state \cite{c3}.

EEG is in the form of time-varying electrical signal measured by small electrodes placed on scalp. Conventional methods of linear signal modeling and analysis, e.g., Fourier transform, auto-regressive model, have been routinely applied to EEG signals in previous research \cite{c4}. Given the enormous complexities of brain dynamics, EEG signal generation is believed to be a highly nonlinear and non-stationary process \cite {c5}. In this study, we investigate the use of Hilbert-Huang transform (HHT) in nonlinear analysis of EEG signals for emotion detection. The HHT algorithm starts with a data-adaptive process of identifying local extrema, which is named Empirical Mode Decomposition (EMD). By EMD, a single-channel EEG signal can be decomposed into a number of intrinsic mode functions (IMFs). By applying Hilbert transform to the IMFs, the instantaneous frequencies of the signal can be obtained \cite{c6}.
In \cite{c7}, EMD was carried out on EEG signals from individual channels independently and Hilbert transform was applied on the decomposed IMFs. EMD may suffer from mode mixing problem 
that violates the local orthogonality of IMFs. To eliminate mode mixing, Ensemble EMD (EEMD) is explored which consists of sifting an ensemble of noise-added signal, whereby averaging of same-index IMFs across ensemble results in EEMD decomposed IMFs \cite{c8}. In \cite{c9}, EEMD was adopted to analyse EEG signal. The Hilbert spectrum provides local non-linear spectral features. These approaches did not consider the spatial relation among the EEG channels. The number of required IMFs may vary across different channels and the exact frequency scale represented by a designated IMF may also be different from one channel to another. In \cite{c10}, it was shown that the inclusion of spatial information in multi-channel EEG could help to improve the performance of emotion classification. This motivates our current investigation on using the multivariate EMD (MEMD) to exploit across-channel information \cite{c8}. The MEMD incorporates a multi-dimensional signal projection method to determine common oscillatory modes of EEG channels. It has been successfully applied to signal analysis tasks that involve multi-sensors \cite{c11}.

Recently, deep learning models have demonstrated technical advantages and performance gain over conventional classifiers like SVM and KNN for EEG based emotion detection.
In \cite{c10, c12, c13, c14}, deep learning models showed better robustness on data samples with high dimensionality and non-linearity. LSTM networks with time-domain features from multi-channel EEG and differential entropy features were investigated in \cite{c13} and \cite{c14} respectively. Using a stacked autoencoder (SAE) with LSTM was proposed to alleviate the linear mixing problem of EEG \cite{c12}. A 2D CNN model was used with temporal and spectral features in \cite{c15, c16}. In \cite{c17}, a hybrid model combining 2D CNN with LSTM was developed to extract spatio-temporal features from raw EEG signals. In CNN, convolution along both temporal and spatial dimensions is expected to facilitate concurrent temporal-spatial feature learning. In \cite{c10, c18, c19}, end-to-end 3D CNN models were applied to extract temporal-spatial features and showed advantages over 2D CNN. 3D CNN with wavelet transform based spectral features was studied in \cite{c20}. In the present study, we investigate a 3D CNN model with the Marginal Hilbert spectrum of all EEG channels as input features for emotion detection.

In the next section, the basics of MEMD for multivariate signals and Hilbert-Huang transform are explained. The 3D CNN based emotion detection system is described in Section 3 and experimental results are reported and discussed in Section 4. Conclusions are stated in Section 5.

\section{SIGNAL PROCESSING METHODS}

\subsection{MEMD and IMF Extraction}
EMD is a data-dependent iterative process by which a uni-variate signal is decomposed into a set of IMFs, which are adaptive with signal variation. IMFs represent the embedded oscillatory modes in the signal. The decomposition method requires extraction of local mean and extremes of the signal. To ensure narrow-band IMFs, the following conditions need to be satisfied.
\begin{itemize}
\item Symmetric upper/lower envelopes, i.e., zero local mean,
\item Number of zero-crossings and extremes are either equal or differ exactly by $1$
\end{itemize}
Applying uni-variate EMD straightforwardly to multi-channel EEG suffers from the non-uniformity and scale alignment problem, i.e., the IMFs of the same index from different channels may correspond to different frequency scales. The multivariate extension of EMD addresses the above problems by generating the same number of scale-aligned IMFs and thus enables coherent time-frequency analysis of all channels. In MEMD, the local mean is determined by taking the average of multidimensional envelopes of a projected signal. The major steps of MEMD are described below \cite{c8}.

Step 1: Given an $n$-variate signal $x(t)$, determine $V$ direction vectors that uniformly sample the $n$-sphere. Let the direction vectors be denoted as $s_{\theta _v}, v=1,...,V$;

Step 2: Project $x(t)$ along each of the direction vectors. The projected signals are denoted as $\left \{ p_{\theta _v}(t) \right \}_{v=1}^{V}$;

Step 3: For each of the projected signals, locate the maxima denoted as $\left \{ t_{\theta _v}^{i} \right \}_{v=1}^{V}$;

Step 4: Interpolate $[t_{\theta _v}^{i}, x\left (t_{\theta _v}^{i} \right )]$ via cubic splines to obtain multivariate envelopes $\left \{ e_{\theta _v}(t) \right \}_{v=1}^{V}$;

Step 5: Compute the mean of the envelopes,
$$
m(t)=\frac{1}{V}\sum_{v=1}^{V}e_{\theta _v}(t)\eqno{(1)};
$$

Step 6: Extract the signal detail as $d(t)=x(t)-m(t)$. If $d(t)$ fulfills the stoppage criterion of multivariate IMF then $d(t)=IMF$, otherwise Let $x(t)=d(t)$ and the whole process is repeated to obtain the detail \cite{c6}.

Step 7: Subtract $d(t)$ from $x(t)$ as $x(t):= x(t)-d(t)$, where $d(t)=IMF$ and return to step 1 to continue the sifting process. Stop shifting when no more extremes can be found after a certain number of iterations and return a monotonic function.

Overall, the MEMD decomposition of signal $x(t)$ can be expressed as
$$
x(t)=\sum_{j=1}^{M}c_j(t)+r(t)\eqno{(2)}
$$
where the n-variate IMFs are denoted as $\left \{ c_j(t) \right \}_{j=1}^{M}$ and $r(t)$ is the residue. The IMFs contain scale-aligned joint rotational modes. 


\subsection{HHT and Marginal Hilbert Spectrum}

The HHT algorithm consists of two parts: MEMD and Hilbert spectral analysis.
For the IMF $c_j(t)$, the Hilbert transform $c_{jH}(t)$ can be obtained as \cite {c6,c21}
$$
c_{jH}(t)=\frac{1}{\pi }P\int_{-\infty}^{\infty}\frac{c_j(\tau )}{t-\tau }d\tau \eqno{(3)}
$$
where $P$ is the Cauchy principal value of the singular integral. The analytic signal of $c_j(t)$ is represented as 
$$
z_j(t)=c_j(t)+ic_{jH}(t)=a_j(t)e^{i\theta_j (t)}\eqno{(4)}
$$
where the instantaneous energy $a_j(t)$ and phase $\theta_j(t)$ can be obtained by
$$
a_j(t)=\sqrt{(c_j(t)^{2}+c_{jH}(t)^{2})}\eqno{(5)}
$$
$$
\theta_j(t)=arctan\left ( \frac{c_{jH}(t)}{c_j(t)} \right )\eqno{(6)}
$$

The instantaneous frequency is obtained as $\omega_j(t)=\frac{d\theta _j(t)}{dt}$. The Hilbert-Huang spectrum, $H_j(\omega ,t)$ is a time-frequency distribution of the signal energy can be represented as follows. 
$$
H_j(\omega ,t)=\left\{\begin{matrix}
a_j(t), \omega =\omega _j(t)\\ 
0, otherwise
\end{matrix}\right.\eqno{(7)}
$$
With $M$ IMFs, $H(\omega ,t)=\sum_{j=1}^{M}H_j(\omega ,t)$.

The Marginal Hilbert spectrum (MHS) is one-dimensional representation of the two-dimensional Hilbert spectrum. It is the integral of Hilbert spectrum over time and represents the power contributed by each frequency component. For a signal frame of length $T$, the MHS can be represented as
$$
h(\omega )=\frac{1}{T}\int_{0}^{T}H(\omega ,t)dt\eqno{(8)}
$$

\section{EMOTION DETECTION SYSTEM DESIGN}

The proposed system for emotion detection comprises two parts: HHT based feature extraction and 3D CNN based classification. They are explained in detail in the following sections. 

\subsection{Feature Extraction and Input Preparation}

The instantaneous spectral information obtained from the Hilbert spectrum of multi-channel EEG are used to constitute the input features for emotion detection. For each EEG channel, the signal is divided into short segments of 1 second long, from each of which the MHS is derived. The MHS indicates the instantaneous spectral energy measured at different frequencies. By aggregating the MHS of signals from all EEG channels, a 2D image map can be constructed. It has the size of $E\times W$, where $E$ denotes the number of channels and $W$ denotes the number of frequency bins on MHS. Subsequently, a 3D feature block is constructed by stacking the 2D image map of 3 consecutive segments, i.e., covering a time interval of 3 seconds. In this way we obtain a spatial-temporal-spectral representation of the input feature of a 3D CNN classifier model, which is to be described in the next sub-section.

Representative spectral features can be obtained from the IMFs that are associated most closely with emotion. Previous research explored that spectral features from high-oscillatory EEG waves were mostly influenced by emotion changes \cite{c22} whereas lower oscillatory waves are dominant for sleep stage \cite{c23}. Different approaches to IMF selection were proposed based on, e.g., correlation coefficients, average power, median frequency and others \cite{c24}. In this work, IMFs with median frequency higher than 4 Hz are considered. The performance of individual IMFs in emotion detection is evaluated to determine their significance.

\subsection{3D CNN Model Architecture}

The convolution kernel of 2D CNN operates in either temporal or spatial dimension whilst 3D kernels perform convolution in both dimensions and extract useful information for classification. The 3D input can be represented by a $E\times W\times M$ feature block. In our proposed system, $E$ represents the $32$ EEG channels, $W=192$ are the spectral components and $M=3$ is the temporal depth covering $3$ consecutive segments. At each convolution layer, the 3D input is convolved with learnable 3D kernel and the 3D output is passed through an activation function. 

\begin{figure}[htb]
\begin{minipage}[b]{1.0\linewidth}
  \centering
  \centerline{\includegraphics[width=6.6cm, height=7.8cm]{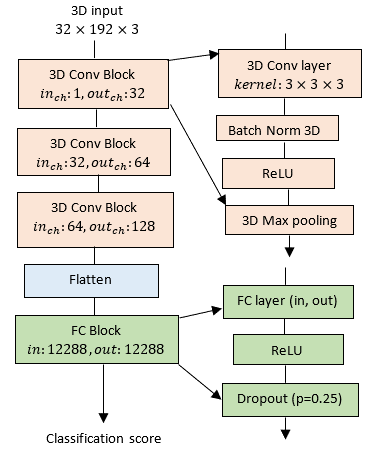}}
\end{minipage}

\caption{3D CNN model architecture for emotion detection}
\label{fig:arch1}
\end{figure}

The adopted CNN model architecture is shown as in Fig. ~\ref{fig:arch1}. It consists of three consecutive convolutional blocks. Each block contains a convolutional layer with $3 \times 3 \times 3$ kernel size which is padded with 1 and stride of 1 to maintain the resolution after convolution. 3D batch normalization is applied followed by rectified linear unit (ReLU) activation function and a 3D max-pooling layer with kernel size of 2. The input channels of the image map for three convolution blocks are 1, 32 and 64 whereas the output channels are 32, 64 and 128 respectively. Flatten operation is applied for transforming into one dimensional feature vector that is the input of a fully connected (FC) block containing FC layer with activation function ReLU and a dropout layer.  

\section{EXPERIMENTS AND RESULTS}

\subsection{EEG Database and Experimental Setup}

The publicly available DEAP database is used in our experiment \cite{c25}. The database contains EEG signals recorded from 32 healthy subjects (age:19-37 yr; gender: 50\% male and 50\% female). For emotion elicitation, each participant was arranged to view 40 video clips. 
A total 1280 multi-channel EEG of one-minute long was recorded during each trial of video viewing. Each participants was asked to give a self-assessment of emotion status after each trial. The assessment is a continuous-value rating score according to the Russell's emotion model \cite{c1}. For both valence and arousal level, the value of self-assessment score is in the range of 1 to 9, which is divided into two classes, namely ``low'' ($<5$) and ``high'' ($>5$). The binary class labels are regarded as the ground-truth emotion labels for model training. 

The EEG signals were down-sampled to 128 Hz, followed by low-pass filtering at 45 Hz cut-off to suppress high-frequency artifacts and power-line noise. Channel re-referencing was performed with the global average scheme. In the binary classification experiments, 5 fold cross validation was applied where $80\%$ of the total samples at video trial level were used for training and  $20\%$ for testing. The models were trained for 200 epochs and outcome was the average accuracy of each fold. The training was conducted with minibatch size of 32 and Adam optimizer where the initial learning rate was 0.0001. A dropout of 0.25 was considered to avoid overfitting.

\subsection{Emotion Detection using 3D CNN Model}

$11$ IMFs are obtained for each channel by applying MEMD. The MHS of an IMF carries pertinent spectral information at a specific frequency range. The first five IMFs are selected according to their median frequency. Table~\ref{tab:table-name1} shows the classification performance attained with individual IMFs computed by MEMD from all EEG channels. It is noted that the high oscillatory mode, i.e., IMF1, is most discriminative with accuracy of $85.37\%$ and $80.56\%$ for binary classification on valence and arousal state respectively. The classification accuracy declines significantly with the lower oscillatory modes. Using IMF4 and IMF5 alone attained $73.18\%$ and $67.27\%$ accuracy on valence state classification and $63.33\%$ and $60.12\%$ on arousal, respectively. On the whole, IMF1$\sim$IMF3 are found to be significant and the lower oscillating intrinsic modes are less important to this task. As IMF1$\sim$IMF3 provide the most discriminative features in an independent manner, spectral features delivered by frequency components ranged in 8-45 Hz are considered for emotion classification in this study.

Table~\ref{tab:table-name2} provides a performance comparison of the proposed system with other state-of-the-art systems of emotion detection reported on the DEAP dataset. 3D CNN with raw EEG as input features in \cite{c19} performs better than LSTM \cite{c12,c13,c14} and 2D CNN \cite{c16} with different types of features. Our proposed system with 3D CNN also outperforms the LSTM and 2D CNN models. Using $3$ dominant IMFs derived by MEMD shows clear advantage over the use of raw EEG. The proposed system achieves accuracy of $89.25\%$ and $86.23\%$ on valence and arousal state detection respectively, which are significantly higher than the reported best results in \cite{c19}.

\begin{table}[h]
\caption{Emotion detection accuracy (\%) for each IMF with 3D CNN}
\label{tab:table-name1}
\begin{center}
\begin{tabular}{|c|c|c c | c c|}
\hline
IMF & Bandwidth & \multicolumn{2}{|c|}{\makebox[0pt]{Valence acc(\%)}} & \multicolumn{2}{|c|}{\makebox[0pt]{Arousal acc(\%)}} \\
 & \textit{(Hz)} & \textit{Train}  & \textit{Test} & \textit{Train} & \textit{Test}\\
\hline
IMF1 & 24-45 & 97.72 & 85.37 & 97.45 & 80.56\\
IMF2 & 13-24 & 96.04 & 84.21 & 97.28 & 76.46\\
IMF3 & 8-13 & 94.03 & 77.25 & 96.43 & 71.73\\
IMF4 & 4-8 & 94.01 & 73.18 & 92.9 & 63.33\\
IMF5 & 2-4 & 90.36 & 67.27 & 91.05 & 60.12\\
\hline
\end{tabular}
\end{center}
\end{table}

\begin{table}[h]
\caption{Performance comparison among recent approaches for subject-independent emotion classification}
\label{tab:table-name2}
\begin{center}
\begin{tabular}{|p{1cm}|p{1.4cm}|p{1.4cm}|p{1.0cm}|p{1.0cm}| }
\hline
Approach & Input & Model & Valence acc(\%)  & Arousal acc(\%)\\
\hline
Ref\cite{c12} & Bandpower & SAE+LSTM & 81.10 & 74.38\\
Ref\cite{c13} & Raw EEG & LSTM & 85.45 & 85.65\\
Ref\cite{c14} & DE & LSTM & 69.06 & 72.97\\
Ref\cite{c16} & Spectrogram & 2D CNN & 80.46 & 76.56\\
Ref\cite{c19} & Raw EEG & 3D CNN & 82.32 & 84.12\\
Proposed & MHS & 3D CNN & 89.25 & 86.23\\
\hline
\end{tabular}
\end{center}
\end{table}

\section{CONCLUSION}

This study has demonstrated the efficacy of applying non-linear analysis technique to multi-channel EEG in characterizing and detecting functional brain activity.
The instantaneous spectral features of high oscillatory modes of multivariate EEG signals are found to be highly effective in the task of emotion detection. The spatial-temporal-spectral feature representations can be effectively modeled by the 3D convolution kernels, showing clear advantages over 2D CNN and LSTM based classifiers.


\end{document}